\documentclass[10pt,twocolumn]{article}
\usepackage[margin=2cm]{geometry}
\usepackage{graphicx}
\usepackage{amsmath,amssymb}
\usepackage{booktabs}
\usepackage{hyperref}
\usepackage{natbib}
\usepackage{xcolor}
\usepackage{caption}
\usepackage[section]{placeins}
\input{numbers_v3.tex}

\title{Evaluation Resolution Confounds Learning-Rule Comparisons\\
in Model--Brain RSA of Early Visual Cortex}

\author{Nils Leutenegger\\
Independent Researcher, Switzerland\\
\texttt{github.com/nilsleut}}

\date{}

\begin{document}
\maketitle

\begin{abstract}
\noindent\textbf{Revision note.} The resolution dependence of the V1 difference between the untrained network and backpropagation reported here is not robust to the analysis choices examined in ``Analysis choices decide whether untrained networks beat trained ones at V1: a multiverse analysis of model--brain RSA'' arXiv:2611.15533. With standardised model units, the slope from 32 to 224 px is +0.002 [$-$0.004, 0.007] instead of +0.031 [0.024, 0.038], and the V1 difference at 224 px is $-$0.011 [$-$0.017, $-$0.005]. Under Euclidean distance the V1 advantage persists without a resolution dependence; in ImageNet models a smaller resolution dependence remained. The backpropagation advantage at LOC holds with standardised units.

\medskip

Representational similarity analysis (RSA) is increasingly used to ask which learning rules give convolutional networks brain-like representations. Because biologically plausible rules such as feedback alignment, predictive coding, and STDP do not scale, studies that include them train small networks on small images (typically $32\times32$ CIFAR) and then compare the trained networks to brain responses recorded for naturalistic stimuli, which are modeled at much higher resolution. We find that a common qualitative result in this setting, that untrained or locally trained networks rival or beat backpropagation at early visual cortex, depends strongly on the resolution at which the network is evaluated. The V1 gap between an untrained network and a backpropagation-trained one widens with evaluation resolution, from $\NV{gap.V1.32}$ (95\% CI $[\NV{boot.V1.32.lo}, \NV{boot.V1.32.hi}]$) at the $32$\,px training resolution to $\NV{boot.V1.224}$ $[\NV{boot.V1.224.lo}, \NV{boot.V1.224.hi}]$ at $224$\,px, growing monotonically across six resolutions ($n=5$ seeds; per-subject RSA on stimulus pairs from different fMRI runs); the gap at the training resolution is $\approx 0$ under the fixed Conv1$\to$V1 mapping and $\NV{bestgap.32}$ under best-layer selection, and the growth with resolution holds under both. The effect is established for the random-versus-backprop contrast, whose two conditions share an initialization, and reproduces in the same direction across three further conditions whose initialization or weight displacement is not matched (feedback alignment, predictive coding, STDP; see Appendix~B). It holds in human fMRI and, directionally, in single-seed macaque electrophysiology, along the full training trajectory, and for two architectures trained at $224$\,px on other data (an ImageNet ResNet-50 and a Swin-Tiny transformer). We test four candidate mechanisms and none accounts for it: train/eval resolution matching, since the ResNet-50 and the transformer also align best at low resolution despite being trained at $224$\,px; low-level Gabor and pixel structure; the normalization state of the untrained baseline, tested with a $2\times2$ calibration design that holds the convolutional weights bit-identical; and convergence of the pooled descriptor toward a global brightness statistic. A fifth experiment, analysed against a criterion fixed before the analysis, separates the two things resolution changes at once. Repeating the sweep on stimuli first reduced to $32$\,px and then upsampled limits the image content to the training resolution while the input still grows. The V1 gap does not shrink: at $224$\,px it is $\NV{s4.V1.up224}$ (95\% CI $[\NV{s4.V1.up224.lo}, \NV{s4.V1.up224.hi}]$) with content limited, against $\NV{boot.V1.224}$ with native content (ratio $\NV{s4.V1.R}$ $[\NV{s4.V1.R.lo}, \NV{s4.V1.R.hi}]$). The gap therefore arises once the input exceeds the training resolution, driven by a decline of backprop; image content above $32$\,px is not needed for it and narrows it. The backprop advantage at LOC, by contrast, disappears under band-limited upsampling. One control result is worth stating separately: a single scalar luminance value per image reaches $\rho = \NV{ref.lum.V1}$ against V1, essentially matching the untrained network's $\NV{rho.rnd.V1.224}$. The first of those numbers concerns the brain data alone and is independent of any model; the second is specific to RSA on globally pooled features, and a fitted readout on the full feature map might place the models higher. What the pair bounds, then, is what this style of comparison can resolve at V1 in this dataset, not model--brain alignment in general. In this setting, the one learning effect that holds across resolution is backprop above untrained, at a higher area (LOC). Comparisons of learning rules or architectures at early visual cortex therefore need to control, and report, the evaluation resolution.
\end{abstract}


\section{Introduction}

A recurring question in NeuroAI is which learning rule gives a network the most brain-like visual representations. The usual approach compares candidate models to neural data with representational similarity analysis (RSA) \citep{kriegeskorte2008}, asking whether backpropagation or a more biologically plausible rule (feedback alignment \citep{lillicrap2016}, predictive coding \citep{rao1999,whittington2017}, or spike-timing-dependent plasticity \citep{bi1998,masquelier2007}) better matches the representational geometry of visual cortex \citep{yamins2016,schrimpf2020}. The biologically plausible rules do not yet scale to large datasets, so any study that includes them trains small networks on small images, typically $32\times32$ CIFAR, and then compares the trained representations against brain responses to naturalistic stimuli modeled at much higher resolution.

One observation has been reported repeatedly: untrained, randomly initialized networks are already quite brain-like, sometimes matching or beating trained networks at early visual areas \citep{saxe2011,truzzi2025}. Our own earlier work reproduced it. Across the five conditions, an untrained network matched or exceeded a backpropagation-trained one at human V1 \citep[at $224$\,px, per subject on cross-run pairs, untrained $\rho = \NV{p1.rnd}$ vs.\ backprop $\rho = \NV{p1.bp}$; $\Delta\rho = \NV{p1.gap}$, 95\% CI {$[\NV{p1.gap.lo}, \NV{p1.gap.hi}]$};][]{leutenegger2026}.

This paper shows that the effect depends strongly on one factor that usually goes uncontrolled: the resolution at which the model is evaluated. Run the same trained models on the brain stimuli across a range of input resolutions and the untrained network's V1 advantage tracks that resolution. It is large when the $32$\,px-trained models are evaluated at $224$\,px, the standard choice, and it vanishes at the $32$\,px training resolution ($\NV{gap.V1.32} \pm \NV{gap.V1.32.sem}$; Fig.~\ref{fig:sweep}, $n=5$ seeds). The ranking one reports at early visual cortex thus rests on an analysis choice that is rarely stated.

This dependence is not specific to our network or to one dataset. It holds across all five conditions, in both human fMRI and macaque electrophysiology, and along the whole training trajectory, where the familiar ``training degrades V1'' result tracks the same dependence. It also holds for two further architectures trained at $224$\,px on different data, an ImageNet ResNet-50 \citep{he2016} and a Swin-Tiny transformer \citep{liu2021}. One candidate is the obvious explanation, a mismatch between training and evaluation resolution. If that were the cause, a model trained at $224$\,px should align best at $224$\,px; instead both the ResNet-50 and the transformer align best at low resolution, exactly as the $32$\,px-trained networks do, across a convolutional and a transformer family alike. A second concerns the baseline itself: the untrained network's batch-normalization layers are at initialization and therefore act as the identity, while every trained condition carries statistics accumulated during training. Calibrating those statistics at the evaluation resolution, with the convolutional weights held bit-identical, leaves the untrained network's V1 advantage essentially intact. The remaining two are low-level accounts. Gabor and pixel structure do not predict V1 alignment across models, and the ResNet-50 is far more Gabor-like than an untrained CNN while aligning less well with V1. Global image statistics come closer, and this is where the sweep turns up its second result: a single scalar luminance value per image reaches $\rho = \NV{ref.lum.V1}$ against the V1 RDM, essentially matching the untrained network's own $\NV{rho.rnd.V1.224}$, and partialling luminance out halves that network's alignment. Luminance similarity also orders the conditions exactly as their resolution slopes do. But it is not the carrier: holding the convolutional weights bit-identical and varying only the normalization statistics separates the two, with one variant lowering its luminance similarity across the sweep while raising its V1 alignment. We report the phenomenon and its boundaries and leave the mechanism open.

Not everything moves with resolution. At a higher area (LOC), backpropagation-trained networks align better than untrained ones at every resolution we tested (LOC backprop$-$untrained $\NV{gap.LOC.32}$ at $32$\,px and $\NV{gap.LOC.224}$ at $224$\,px, $5/5$ seeds at both). Learning does reshape representations; what it does at early visual cortex is simply swamped by the evaluation resolution.

\textbf{Contributions.} (1)~We identify and quantify an evaluation-resolution dependence in model--brain RSA at early visual cortex, and show it is general across conditions, two species, the training trajectory, and three architecture families. (2)~We test four candidate mechanisms and rule out all four, three of them by interventions that hold the convolutional weights bit-identical. (3)~We separate the two things evaluation resolution changes at once and show that the V1 gap does not require image content above the training resolution: it arises once the input exceeds that resolution, driven by a decline of backprop. (4)~We show that a single scalar luminance value per image matches the untrained network's V1 alignment in this dataset, and that luminance similarity orders the conditions exactly as their resolution slopes do without carrying the effect, which bounds what these comparisons can resolve. (5)~We isolate a learning effect that does survive across resolution, at LOC, and we recommend evaluating alignment at the training resolution and at several others.

\section{Methods}

\textbf{Learning rules and architecture.} We compare five conditions: random (untrained) weights, backpropagation (BP), feedback alignment (FA), predictive coding (PC), and STDP. All use the same small convolutional architecture (three convolutional blocks Conv1--Conv3, each $3\times3$ kernels with batch normalization, ReLU and $2\times2$ max-pooling; $32/64/128$ filters, followed by FC1 of $512$ units and a $10$-way head FC2). Each condition is trained on an $8{,}000$-image subset of CIFAR-10 at $32\times32$ resolution, batch size $128$, for $40$ epochs, with $5$ random seeds ($42, 123, 456, 789, 1337$). Optimizers and rates differ by rule: BP uses Adam (lr $10^{-3}$, weight decay $10^{-4}$, cosine schedule, gradient clipping at $1.0$, dropout $0.3$, cross-entropy); FA uses SGD (lr $5\times10^{-4}$, momentum $0.9$) with fixed random feedback weights; PC refines representations for $T=10$ inference steps (inference rate $0.02$) and updates feedforward weights by local prediction error (lr $10^{-4}$) with an Adam-trained readout; STDP converts activations to Poisson spike trains ($10$ timesteps) and updates convolutional weights by a spike-timing kernel ($A_\pm=0.003$, $\tau_\pm=20$\,ms, lr $5\times10^{-4}$) with an Adam-trained readout. Full details follow \citet{leutenegger2026}.

\textbf{Normalization controls.} The untrained condition's batch-normalization layers are at initialization. To test whether that difference drives the results, we add calibrated variants of it: starting from the untrained network, we estimate batch-normalization statistics from $50$ forward passes in training mode with no weight updates, using an unbiased cumulative average rather than the default exponential moving average, and a deterministic loader with no augmentation. The calibration set is either CIFAR-10 or THINGS images drawn from a pool of $1{,}134$ concepts that contribute no evaluation stimulus, and the calibration resolution is either $32$\,px or the resolution at which the model is subsequently evaluated, giving a $2\times2$ design. As a check on the calibration procedure itself we additionally calibrate on the $720$ evaluation stimuli. After every calibration we verify that all convolutional weights are bit-identical to the plain untrained condition.

\textbf{Additional architectures (eval-only).} To test architecture-independence we add two standard ImageNet-pretrained models, evaluated without further training: a ResNet-50 \citep[torchvision \texttt{IMAGENET1K\_V2};][]{he2016} and a Swin-Tiny transformer \citep[torchvision \texttt{IMAGENET1K\_V1};][]{liu2021}.

\textbf{Brain data.} Human: THINGS-fMRI representational dissimilarity matrices (RDMs) for V1, V2, LOC and IT over $720$ object images from THINGS-fMRI \citep[3T Siemens Magnetom Prisma, $3$ subjects;][]{hebart2023}; each model RDM is compared with each subject's RDM separately. Macaque: electrophysiology via Brain-Score, namely FreemanZiemba2013 \citep[V1, V2; $135$ texture stimuli;][]{freeman2013} and MajajHong2015 \citep[V4, IT; $3{,}200$ object presentations;][]{majaj2015}.

\textbf{Stimulus preprocessing.} Stimuli are resized with bilinear interpolation on the PIL image (\texttt{transforms.Resize(px)} followed by \texttt{CenterCrop(px)}), which scales the shorter edge and preserves aspect ratio, so the retained image region is identical at every evaluation resolution; downsampling is antialiased. Inputs are normalized with each model's own training statistics: CIFAR-10 channel statistics for the custom CNN, ImageNet statistics for the ResNet-50 and the Swin-Tiny. All models are placed in evaluation mode for feature extraction, so batch normalization uses its stored running statistics and these are not updated by the evaluation pass; this is asserted at extraction time for every condition. Note that the untrained network's batch-normalization layers retain their initialized running statistics (mean $0$, variance $1$), so normalization is close to the identity for that condition, whereas the trained networks carry statistics accumulated at the $32$\,px training resolution.

\textbf{RSA.} For each model we extract layer activations for every stimulus, global-average-pool convolutional or stage feature maps, and build a model RDM with correlation distance. For the predictive-coding condition, features are the representations produced by the inference loop ($T=10$ steps at inference rate $0.02$), not a single feedforward pass; using the feedforward pass instead gives a different model. Model and brain RDMs are compared by Spearman correlation. Values in the text are the mean $\pm$ SEM across the $5$ seeds, with the number of seeds in which a difference is positive, in the per-subject convention: each model RDM is correlated with each subject's RDM and the three correlations are averaged, using only the $210{,}205$ of $258{,}840$ stimulus pairs whose two stimuli lie in different fMRI runs for all three subjects (two subjects saw the stimuli in identical run order, so same-run pairs share a run component). The headline V1 gap additionally carries a 95\% stimulus-bootstrap CI ($1{,}000$ resamples of the $720$ stimuli, with the cross-run pair set recomputed in each). The figures use the same convention; their error bars are 95\% stimulus-bootstrap CIs of the seed mean (Fig.~\ref{fig:arch}: of the single ResNet-50 and Swin-Tiny checkpoints), and Fig.~\ref{fig:gabor} shows single-seed points as in v2. The model set is the $30$-checkpoint retrain of the repaired sweep; the endpoint study \citep{leutenegger2026} uses a nominally identical retrain of the same five-seed configuration, and its gap at the headline cell ($\NV{p1.gap}$) differs from ours ($\NV{gap.V1.224}$) by less than the seed spread. The conventions give different absolute values for the same models and the same pattern: the V1 Random$-$Backprop gap runs from $\NV{gap.V1.32}$ ($\NV{gap.V1.32.pos}/5$ seeds) at $32$\,px to $\NV{gap.V1.224}$ ($\NV{gap.V1.224.pos}/5$) at $224$\,px per subject on cross-run pairs, from $\NV{mr.cr.V1.32}$ to $\NV{mr.cr.V1.224}$ with RDM averaging on cross-run pairs, and from $\NV{mr.all.V1.32}$ to $\NV{mr.all.V1.224}$ with RDM averaging on all pairs. Differences across seeds, such as the Random$-$Backprop gap, are tested with a paired, one-sided sign-flip permutation test over the five seeds ($2^5=32$ assignments; smallest attainable $p = 1/32 \approx 0.031$), following the companion training-dynamics study. All differences, factor effects and standard errors are computed at full precision; they may therefore differ in the last digit from arithmetic on the rounded values printed. Layer$\to$ROI mapping for the custom CNN: Conv1$\to$V1/V2, Conv2$\to$V1/V2, Conv3$\to$LOC, FC1$\to$IT --- six layer--ROI pairs. The headline V1 result uses Conv1 throughout; Conv2 enters the best-layer analysis of Appendix~C. For ResNet-50: \texttt{layer1}$\to$V1/V2, \texttt{layer2}$\to$V4, \texttt{layer4}$\to$IT. For Swin-Tiny: early stage$\to$V1/V2, late stage$\to$IT.

\textbf{Interpreting the scale.} Absolute Spearman $\rho$ values in this setting are low. The $720$ evaluation stimuli are single-presentation, so no within-subject reliability can be estimated on them. With three subjects the informative reference is the leave-one-subject-out lower bound of \citet{nili2014}, each subject's RDM against the mean of the other two: on cross-run pairs it is $\NV{lb.V1}$ (95\% CI $[\NV{lb.V1.lo}, \NV{lb.V1.hi}]$) at V1, $\NV{lb.V2}$ at V2, $\NV{lb.LOC}$ at LOC and $\NV{lb.IT}$ at IT \citep{leutenegger2026}. The corresponding upper bound is not informative at $N=3$, where each subject contributes a third of its own target. The ceiling figure of earlier versions of our endpoint study, described there as split-half reliability, was not reproducible and is not used.

The bound that does hold requires no ceiling: a single scalar luminance value per image reaches $\rho = \NV{ref.lum.V1}$ against the V1 RDM, essentially matching the untrained network at $224$\,px ($\NV{rho.rnd.V1.224}$) and the lower bound at V1. That number involves no model and no readout, so it bounds the brain side directly. Occupying most of the resolvable signal at V1 under this comparison is a low bar, and none of our conditions clears it by much. What makes the differences we report interpretable is their consistency across seeds and their monotone behaviour across the sweep, not their absolute size.

\textbf{Resolution sweep.} This is the central manipulation. We evaluate each already-trained model on the brain stimuli at six input resolutions ($32, 64, 96, 128, 160, 224$\,px), with the network, weights, and normalization held fixed. Training is always at $32$\,px; only the evaluation resolution varies, so the sweep adds no retraining.

\textbf{Content control.} To separate the number of pooled positions from the image detail available at each, we repeat the entire sweep in a second arm in which every stimulus is first resized to $32$\,px and then upsampled to the evaluation resolution, with identical interpolation, antialiasing, cropping and normalization in both arms. Both arms are evaluated on the same model object in the same process, so the contrast between them is exact. The nondeterministic kernel is convolution \emph{weight}-gradient backward. Conditions whose stored parameters are updated from convolution weight gradients (backprop, feedback alignment) reproduce across separate runs only to about $2\times10^{-3}$ at V1, with a maximum of $5\times10^{-3}$ across all layer--ROI cells; conditions whose convolutional weights are not so updated reproduce bit-identically. This includes STDP, which does backpropagate through the convolutions to train its batch-normalization parameters but discards the convolutional weight gradient and sets those weights by a local rule. We confirmed the source with a paired probe: two runs under deterministic algorithms are bit-identical, two default runs are not.

\textbf{Low-level controls.} To test low-level-statistics accounts we build, at each resolution, a Gabor filterbank RDM (a parameter-free V1-like model: energy of oriented Gabor filters at four orientations $\times$ three spatial frequencies) and a raw-luminance (pixel) RDM, and correlate each with the model RDMs and with the V1 brain RDM. To test whether pooling converges the descriptor toward a global image statistic we add three further parameter-free references over the same stimuli: mean RGB (three dimensions per image), a joint RGB histogram ($8\times8\times8$ bins), and mean luminance (one dimension per image), each with correlation distance except mean luminance, which uses absolute difference. These are resolution-invariant by construction; we verify this (RDMs built at $32$ versus $224$\,px correlate at $\rho \geq 0.957$), so any resolution trend in their similarity to a model comes from the model side. Partial RSA residualizes both model and brain RDM vectors against a reference by rank-based linear regression.

\section{Results}

The structure of what follows is an elimination followed by a localization. Section~3.1 establishes that the V1 ranking moves with evaluation resolution. The three sections after it try to explain that dependence away and none succeeds: the untrained baseline's missing normalization (\S3.2), a penalty for evaluating a model away from its training resolution (\S3.3), and low-level image statistics, including the convergence of the pooled descriptor toward a global brightness measure (\S3.4). Section~3.5 then separates the two things resolution changes at once and finds that the V1 gap follows the input size, not the image content. Sections~3.6 and 3.7 show that it also governs the training trajectory and the macaque data, and identify the one comparison it does not touch.

\subsection{The V1 rule ranking is a function of evaluation resolution}

\begin{figure}[tbp]
\centering
\includegraphics[width=\columnwidth]{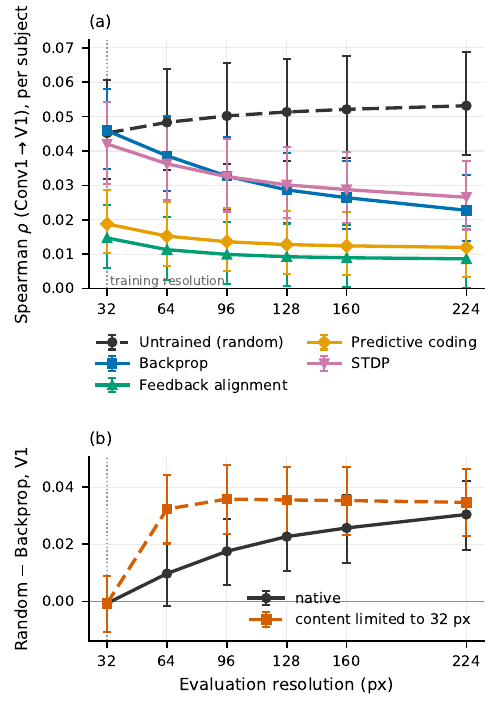}
\caption{\textbf{V1 alignment is a function of evaluation resolution.} (a)~Mean Spearman $\rho$ (Conv1$\to$V1, per subject, cross-run pairs) across $5$ seeds, with 95\% stimulus-bootstrap CIs. Every trained condition peaks at the $32$\,px training resolution and declines; only the untrained random network rises. The Random$-$Backprop gap is $\approx 0$ at $32$\,px and grows to $\NV{gap.V1.224}$ at $224$\,px; the same pattern holds, shifted downward, after low-level statistics are partialled out (\S3.4). (b)~Random$-$Backprop gap, paired per seed and averaged over the $5$ seeds, with paired 95\% stimulus-bootstrap CIs, for native stimuli and for stimuli with content limited to $32$\,px (\S3.5).}
\label{fig:sweep}
\end{figure}

Across the sweep, the early-visual rule ranking is not fixed; it shifts with evaluation resolution (Fig.~\ref{fig:sweep}). Every trained condition aligns best at or near the $32$\,px training resolution and falls off as resolution rises: backprop from $\rho=\NV{rho.bp.V1.32}$ at $32$\,px to $\rho=\NV{rho.bp.V1.224}$ at $224$\,px, feedback alignment from $\NV{rho.fa.V1.32}$ to $\NV{rho.fa.V1.224}$, predictive coding from $\NV{rho.pc.V1.32}$ to $\NV{rho.pc.V1.224}$, STDP from $\NV{rho.stdp.V1.32}$ to $\NV{rho.stdp.V1.224}$ ($5$ seeds). The untrained network goes the other way, climbing from $\rho=\NV{rho.rnd.V1.32}$ to $\rho=\NV{rho.rnd.V1.224}$.

The gap between untrained and backprop at V1, which is what ``untrained $\geq$ trained'' claims rest on, follows directly from this. It runs from $\NV{gap.V1.32} \pm \NV{gap.V1.32.sem}$ at the $32$\,px training resolution ($\NV{gap.V1.32.pos}/5$ seeds positive, not significant; 95\% CI $[\NV{boot.V1.32.lo}, \NV{boot.V1.32.hi}]$) to $\NV{gap.V1.224} \pm \NV{gap.V1.224.sem}$ at $224$\,px ($\NV{gap.V1.224.pos}/5$ seeds positive; 95\% CI $[\NV{boot.V1.224.lo}, \NV{boot.V1.224.hi}]$), growing monotonically across the sweep. With five seeds the sign-flip test has a floor of $p = 1/32 \approx 0.031$; the dose--response relationship across six resolutions, not the $p$-value at any one of them, is what the claim rests on. At the resolution the models were trained on there is no effect; at the resolution normally used to evaluate them it is large.

The fixed layer-to-ROI mapping does not create this. Selecting, for each condition and each resolution, whichever layer aligns best with V1 leaves the best layer unchanged across the sweep for every condition but predictive coding, and the gap grows in the same way and further: $\NV{bestgap.32} \pm \NV{bestgap.32.sem}$ ($\NV{bestgap.32.pos}/5$ seeds) at $32$\,px to $\NV{bestgap.224} \pm \NV{bestgap.224.sem}$ ($\NV{bestgap.224.pos}/5$) at $224$\,px (Appendix~C). One thing does change: under best-layer selection the gap at the training resolution is no longer $\approx 0$ but $\NV{bestgap.32}$. That the gap \emph{vanishes} at $32$\,px is a property of the fixed Conv1 mapping; that it \emph{grows} with resolution is not. The selection is also informative in its own right: the untrained network's best V1 layer is Conv2 while backprop's is Conv1, so a fixed Conv1$\to$V1 mapping is not neutral between the two conditions being compared.

\subsection{It is not the untrained baseline's normalization}

A candidate that concerns the baseline rather than the models. The untrained condition is a network whose batch-normalization layers have never seen data: running mean $0$, running variance $1$, no batches tracked, so in evaluation mode they are the identity. The untrained baseline is therefore not a network with matched normalization but a network with none, while every trained condition carries statistics accumulated at $32$\,px. That difference is correlated with the contrast the comparison is meant to isolate, so in principle it could produce the whole effect.

\begin{figure}[tbp]
\centering
\includegraphics[width=\columnwidth]{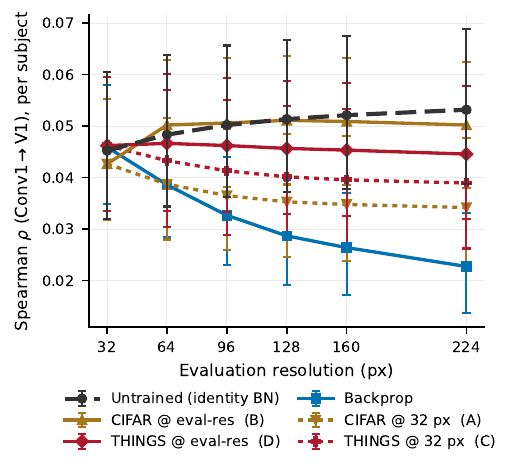}
\caption{\textbf{The effect is not the untrained baseline's normalization.} V1 alignment vs.\ evaluation resolution for the untrained network with its initialized (identity) batch-normalization statistics and for four calibrated variants ($2\times2$: CIFAR-10 or disjoint THINGS images, at $32$\,px or at the evaluation resolution), with the backprop-trained network for reference. Calibration leaves the convolutional weights bit-identical. Resolution-matched calibration costs little and preserves the untrained network's V1 advantage over backprop.}
\label{fig:bncal}
\end{figure}

It does not (Fig.~\ref{fig:bncal}). Calibrating at the evaluation resolution costs little: $\NV{cal.b-random.224} \pm \NV{cal.b-random.224.sem}$ at $224$\,px with CIFAR-10 statistics ($\NV{cal.b-random.224.pos}/5$ seeds positive) and $\NV{cal.d-random.224} \pm \NV{cal.d-random.224.sem}$ with THINGS statistics from a disjoint image pool ($\NV{cal.d-random.224.pos}/5$). Calibrating at $32$\,px and evaluating higher costs more ($\NV{cal.a-random.224} \pm \NV{cal.a-random.224.sem}$ and $\NV{cal.c-random.224} \pm \NV{cal.c-random.224.sem}$ at $224$\,px). The $2\times2$ design separates the two factors: matching the calibration resolution is worth $\NV{cal.B-A.224} \pm \NV{cal.B-A.224.sem}$ for CIFAR and $\NV{cal.D-C.224} \pm \NV{cal.D-C.224.sem}$ for THINGS ($5/5$ seeds each), whereas changing the calibration set at matched resolution is worth $\NV{cal.D-B.224} \pm \NV{cal.D-B.224.sem}$ ($\NV{cal.D-B.224.pos}/5$). What the calibration costs is itself mostly a resolution effect, not a distribution effect.

The untrained advantage survives all of it. Against a resolution-matched calibrated baseline the untrained network still exceeds backprop at V1 at $224$\,px by $\NV{cal.b-bp.224}$ (CIFAR calibration) or $\NV{cal.d-bp.224}$ (THINGS calibration), $5/5$ seeds in both cases, against $\NV{gap.V1.224}$ for the uncalibrated baseline; and the gap remains $\approx 0$ at $32$\,px under every variant. The resolution dependence is not an artifact of an unnormalized baseline. We report seed counts rather than standard errors for these paired contrasts because at $n=5$ the paired SEM is itself unstable: it ranges from $\NV{cal.sem.min}$ to $\NV{cal.sem.max}$ across variants whose per-seed differences are comparable, depending on how strongly a given variant happens to covary with backprop across the five initializations.

The natural reading of what calibration does cost, that alignment tracks how well the stored statistics match the activations the evaluation stimuli actually produce, is contradicted by the same experiment. Calibrating on the $720$ evaluation images gives, by construction, an almost exact estimate of those activations (mean gap $0.0003$, variance ratio $1.00$ at the first block) and is the worst of the eight variants, costing $-0.030 \pm 0.006$ relative to the identity baseline ($0/5$ seeds positive; v2 convention, RDM averaged over subjects on all pairs: the RDMs of this probe were not stored, so it is the one value not recomputed per subject on cross-run pairs). The identity baseline's stored variance is wrong by a factor of $7$ at the first block, and by $37$ and $202$ at the second and third, and it aligns best of all. Since the Conv1$\to$V1 comparison depends on the first block alone, this is not a downstream effect. We can therefore say what the calibration cost is not, but not what it is, and we leave that open as well.

\subsection{Train/eval resolution matching does not account for it: ResNet-50 and Swin-Tiny}

\begin{figure}[tbp]
\centering
\includegraphics[width=\columnwidth]{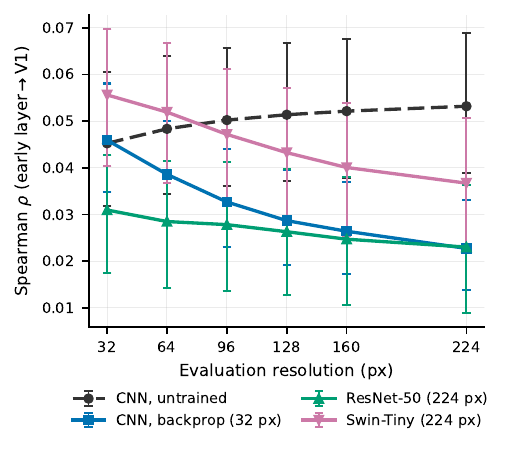}
\caption{\textbf{Architecture-independent, not train/eval matching.} V1 alignment vs.\ evaluation resolution for three architecture families. All three trained models, including ResNet-50 and Swin-Tiny (both trained at $224$\,px), fall toward $224$\,px; only the untrained CNN rises. There is no crossover at the $224$\,px training resolution.}
\label{fig:arch}
\end{figure}

The obvious explanation is that a model is penalized when evaluated away from its training resolution, which predicts that a model trained at $224$\,px should peak at $224$\,px. It does not. An ImageNet-trained ResNet-50 and a Swin-Tiny transformer, both trained at $224$\,px, show V1 alignment that falls toward $224$\,px and peaks at low resolution, just as the $32$\,px-trained networks do (Fig.~\ref{fig:arch}): ResNet-50 from $\rho=\NV{arch.resnet50.32}$ to $\rho=\NV{arch.resnet50.224}$, Swin-Tiny from $\rho=\NV{arch.swin_t.32}$ to $\rho=\NV{arch.swin_t.224}$. The pattern thus holds across three architecture families (custom CNN, ResNet-50, Swin-Tiny) and does not depend on training resolution. Only the untrained network climbs with resolution.

\subsection{Low-level accounts: Gabor, pixel, and global colour}

\begin{figure}[tbp]
\centering
\includegraphics[width=\columnwidth]{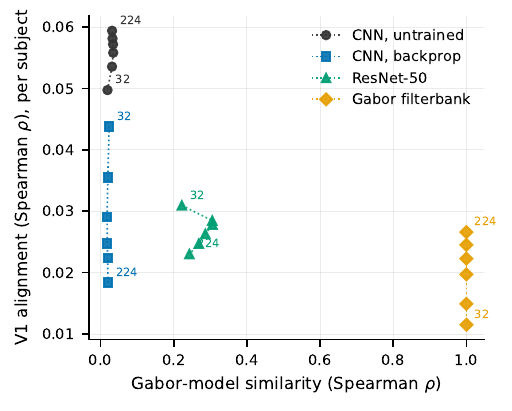}
\caption{\textbf{The effect is not a simple Gabor account.} Gabor-model similarity (x) vs.\ V1 alignment (y) across models and resolutions. We see no relationship between the two; the ResNet-50, far more Gabor-like, is less V1-aligned than the untrained CNN.}
\label{fig:gabor}
\end{figure}

A candidate account is that untrained filters look V1-like because they capture low-level image statistics. The Gabor version fails too (Fig.~\ref{fig:gabor}). A Gabor filterbank is itself only weakly related to V1 ($\rho\approx\NV{gabor.V1.min}$--$\NV{gabor.V1.max}$), and across models and resolutions we see no relationship between Gabor-similarity and V1 alignment, though with three model families the analysis is not powered to exclude a weak one. The clearer evidence is the counterexample: the ResNet-50's early stage is about ten times more Gabor-like than the untrained CNN (Gabor-similarity $\approx0.22$--$0.31$ vs.\ $\approx0.02$--$0.03$) and yet aligns less well with V1. Being Gabor-like does not buy V1 alignment here.

A raw-luminance (pixel) RDM behaves the same way. It is itself weakly related to V1 ($\rho=\NV{ref.pixel.V1}$), and the models are barely pixel-like, with no ordering that tracks V1 alignment. The endpoint study reached the same conclusion by a different route, partialling a pixel RDM out of both model and brain RDMs and finding the V1 ordering fully preserved, with decreases of $\NV{p1.pr.min}$--$\NV{p1.pr.max}$ \citep{leutenegger2026}; our own partial-RSA implementation reproduces that band ($\NV{unt.pixel_decrease}$ at $224$\,px) before being applied to the references below. V1-like filter structure is, after all, exactly what unsupervised learning on natural image statistics is known to produce \citep{olshausen1996}, so a low-level account is the natural first suspect.

A further low-level account concerns the pooling itself. Global average pooling over the Conv1 map covers $256$ spatial positions at $32$\,px and $12{,}544$ at $224$\,px, a $49$-fold increase, so the pooled descriptor converges toward a global image statistic as resolution rises; with random filters that limit is close to a global colour or luminance measure. This is the account that comes closest to working, and the one whose failure is most informative. Global statistics are a large part of why the untrained network aligns with V1 at all: a single scalar luminance value per image reaches $\rho = \NV{ref.lum.V1}$ against the V1 RDM, essentially matching the untrained network's own $\NV{rho.rnd.V1.224}$ at $224$\,px, and partialling luminance out of both model and brain RDMs halves the untrained network's alignment ($\NV{rho.rnd.V1.224} \to \NV{unt.partial_lum}$). That early visual cortex is strongly driven by low-level image properties is not itself new \citep{goddard2020}; what we quantify is that in this dataset a one-dimensional descriptor matches the best model we tested under this comparison. The luminance figure itself involves no model and no readout, so it bounds the brain side directly; the models' figures are specific to rank correlation on globally pooled features, and a fitted readout on the full feature map could well place them higher. The claim is therefore about what RSA on pooled features can resolve at V1 here, not about model--brain alignment in general. Mean chromaticity fails in the opposite direction: among the five learning-rule conditions the untrained network is the \emph{least} mean-RGB-like ($0.24$) and backprop the most ($0.54$).

Which reference one uses decides the verdict. Partialling all four low-level references jointly out of both model and brain RDMs leaves the untrained$-$backprop gap climbing from $\NV{gap.joint4.32}$ at $32$\,px to $\NV{gap.joint4.224}$ at $224$\,px, against $\NV{gap.V1.224}$ uncorrected, so the resolution dependence is not simply a colour effect being read twice; partialling mean luminance alone cuts the gap at $224$\,px from $\NV{gap.V1.224}$ to $\NV{gap.lum.224}$ without flattening it. Convergence toward the colour histogram does not order the conditions in the way the account requires: the plain and calibrated untrained conditions converge at nearly the same rate ($+0.048$ and $+0.052$) while their V1 slopes have opposite signs. Convergence toward mean luminance does, and more sharply than any other reference. Across the six conditions the change in luminance similarity over the sweep and the change in V1 alignment rank together at Spearman $\rho = \NV{six.order}$, and across five conditions that share bit-identical convolutional weights and identical pooling geometry, differing only in their stored normalization statistics, the ordering is perfect ($\rho = 1.00$, exact permutation $p = 0.017$; $\rho = \NV{within.rho25}$ over all $25$ variant-seed points). Within those five, the corresponding values for mean RGB and pixel similarity are $-0.30$ and $-0.70$, so what is ordered is overall brightness specifically, not colour or layout. Mean luminance is also the reference that best predicts V1 in the first place ($\rho = \NV{ref.lum.V1}$ against $\NV{ref.hist.V1}$, $\NV{ref.rgb.V1}$ and $\NV{ref.pixel.V1}$ for colour histogram, mean RGB and pixels).

The ordering is nevertheless not a pooling mechanism, and the same experiment shows why. Under that account, averaging over $49$ times more positions should drive the descriptor toward the global statistic whatever the normalization, since all five conditions share the filters and the pooling. Only the unnormalized network does so ($+0.015$, $5/5$ seeds); the four calibrated variants move \emph{away} from luminance over the same sweep ($-0.011$ to $-0.046$). One variant separates the two quantities outright: calibrating on CIFAR-10 at the evaluation resolution lowers luminance similarity across the sweep ($-0.011$, $1/5$ seeds positive) while raising V1 alignment ($\NV{within.B.dV1}$, $\NV{within.B.dV1.pos}/5$), so the response rises as the putative dose falls. And within a variant, across seeds, with the normalization state held fixed, the two quantities do not covary consistently (correlations from $-0.20$ to $+0.90$). The relationship lives between normalization states rather than along the pooling axis, which makes it a shared dependence rather than a chain from pooling to alignment.

We therefore report this account as tested and not supported. What survives it is an observation rather than a mechanism: whatever moves V1 alignment in this setting also moves similarity to a single brightness scalar, and we can explain neither.

\subsection{Input size, not image content}

\begin{figure*}[tp]
\centering
\includegraphics[width=\textwidth]{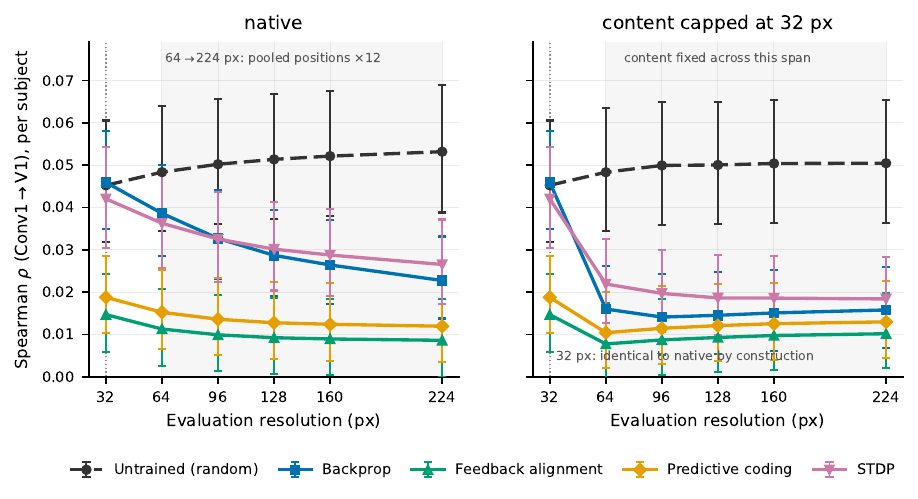}
\caption{\textbf{With image content limited to the training resolution, the V1 gap is as large as with native content.} V1 alignment (Conv1$\to$V1) vs.\ evaluation resolution, mean over $5$ seeds with 95\% stimulus-bootstrap CIs (per subject, cross-run pairs), with stimuli at their native resolution (left) and with their content capped at $32$\,px before upsampling (right). The arms are identical at $32$\,px by construction. What differs is the shape. At native resolution the trained conditions decline steadily across the whole sweep; with content limited they drop at the first step, as soon as the input exceeds the training resolution, and then run flat. Across the shaded $64\to224$\,px span the pooled positions increase twelvefold while the right-hand arm's content stays fixed, and the Random$-$Backprop gap opens by $\NV{ups.gapopen.UPSAMPLED}$ there against $\NV{ups.gapopen.NATIVE}$ on the left. Only the untrained network rises in both arms.}
\label{fig:content}
\end{figure*}

Varying evaluation resolution varies two things at once: the number of positions the descriptor is pooled over, and the image detail available at each. A final experiment separates them (Fig.~\ref{fig:content}). We repeat the sweep in a second arm in which every stimulus is first reduced to $32$\,px and then upsampled to the evaluation resolution, so the information content is capped at the training resolution while the network still pools over the full number of positions. At $32$\,px the second resize is a no-op and the two arms are identical by construction, which we verify (max $|\Delta\rho| = 0$ across all cells).

We evaluated the comparison against a criterion fixed before the analysis (October 2026): the ratio $R = \mathrm{Gap}_{\mathrm{UP}}(224)/\mathrm{Gap}_{\mathrm{NAT}}(224)$ of the V1 Random$-$Backprop gap with content limited to the gap with native content, with a stimulus-bootstrap CI on resamples shared by both arms. $R \geq 0.5$ with a CI of $\mathrm{Gap}_{\mathrm{UP}}$ excluding $0$ was to mean that the effect depends on input size; a CI of $R$ including $0$, that it depends on image content. The result is the first case. At $224$\,px the gap is $\NV{s4.V1.up224}$ $[\NV{s4.V1.up224.lo}, \NV{s4.V1.up224.hi}]$ with content limited and $\NV{boot.V1.224}$ $[\NV{boot.V1.224.lo}, \NV{boot.V1.224.hi}]$ with native content, $R = \NV{s4.V1.R}$ $[\NV{s4.V1.R.lo}, \NV{s4.V1.R.hi}]$; limiting the content widens the gap by $\NV{s4.V1.diff}$ $[\NV{s4.V1.diff.lo}, \NV{s4.V1.diff.hi}]$.

The shape shows where the gap comes from. With content limited it appears in full at the first step ($\NV{s4.V1.up64}$ at $64$\,px) and changes little thereafter: it opens by $\NV{ups.gapopen.UPSAMPLED}$ from $64$ to $224$\,px, against $\NV{ups.gapopen.NATIVE}$ with native content, so growing the number of pooled positions from $1{,}024$ to $12{,}544$ does not carry it either. It is carried by backprop. With content limited, backprop's V1 alignment falls from $\NV{rho.bp.V1.32}$ at $32$\,px to $\NV{s4.bp.up.min}$--$\NV{s4.bp.up.max}$ at every larger input size; with native content it declines gradually to $\NV{rho.bp.V1.224}$ at $224$\,px. The untrained network stays between $\NV{rho.rnd.V1.32}$ and $\NV{rho.rnd.V1.224}$ in both arms. Image content above the training resolution is therefore not needed for the V1 gap; where it is present, it keeps backprop's alignment higher and narrows the gap. The gap arises once the input exceeds the training resolution.

LOC behaves differently. There the backprop advantage is invariant to resolution with native content ($\NV{gap.LOC.32}$ at $32$\,px, $\NV{gap.LOC.224}$ at $224$\,px) but disappears with content limited ($\NV{s4.LOC.up224}$ $[\NV{s4.LOC.up224.lo}, \NV{s4.LOC.up224.hi}]$ at $224$\,px; difference from native $\NV{s4.LOC.diff}$ $[\NV{s4.LOC.diff.lo}, \NV{s4.LOC.diff.hi}]$): it needs natural broadband image content. The ratio $R$ is not informative at LOC ($\NV{s4.LOC.R}$ $[\NV{s4.LOC.R.lo}, \NV{s4.LOC.R.hi}]$) because its denominator is close to zero.

One caution applies to both areas. Upsampled images are smoother than natural images of the same size; the comparison separates information content from input size, but it does not isolate which property of the image is responsible for backprop's decline at V1 or for the loss of its advantage at LOC.

\subsection{It generalizes over training time and across species}

\begin{figure*}[tp]
\centering
\includegraphics[width=\textwidth]{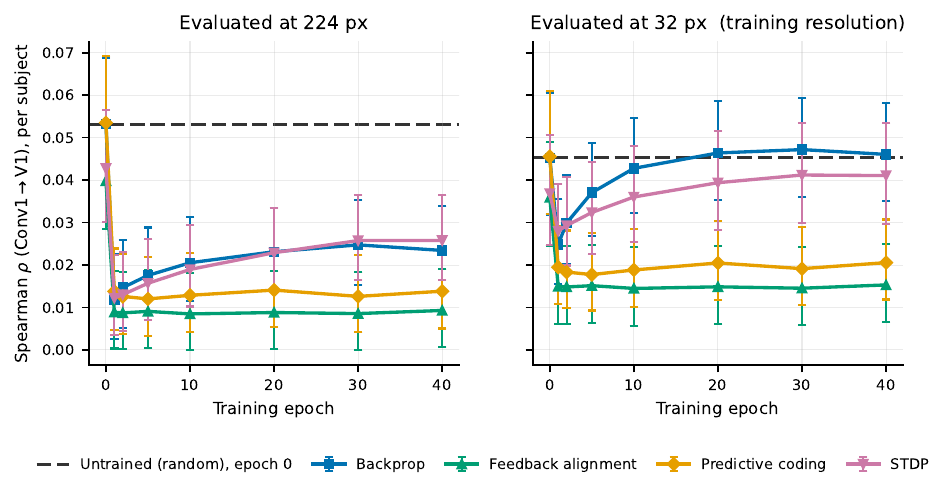}
\caption{\textbf{``Training degrades V1'' tracks the same dependence.} V1 alignment (Conv1$\to$V1) over training, evaluated at $224$\,px (left) and at the $32$\,px training resolution (right), mean over seeds of per-subject $\rho$ on cross-run pairs, with 95\% stimulus-bootstrap CIs. At $224$\,px backprop drops in the first epoch and recovers only partly; at $32$\,px it returns to baseline, showing no net degradation.}
\label{fig:dynamics}
\end{figure*}

The same thing happens over training, not just at the endpoint (Fig.~\ref{fig:dynamics}). At $224$\,px backprop V1 alignment drops sharply in the first epoch, from $\NV{td.bp224.e0}$ to $\NV{td.bp224.e1}$, and then recovers partly, to $\NV{td.bp224.e40}$ at epoch $40$ (net change $\NV{td.bp.224} \pm \NV{td.bp.224.sem}$, $\NV{td.bp.224.neg}/5$ seeds negative), which is the published ``training degrades early-visual alignment'' result; evaluated at the $32$\,px training resolution the same run returns to its starting value ($\NV{td.bp.32} \pm \NV{td.bp.32.sem}$, $\NV{td.bp.32.neg}/5$ seeds negative). For backpropagation, what looked like degradation was the resolution gap widening as training proceeded. This is not uniform across rules: predictive coding and feedback alignment lose alignment at the training resolution as well ($\NV{td.pc.32} \pm \NV{td.pc.32.sem}$ and $\NV{td.fa.32} \pm \NV{td.fa.32.sem}$; $5/5$ and $4/5$ seeds negative), so for those two the loss is not purely a resolution artifact. For predictive coding the loss is also unlikely to be a weight effect: over the same $40$ epochs that condition displaces its first convolutional layer by about $2\%$ of the initialization norm (Appendix~B), which is hard to reconcile with an alignment change of $0.03$--$0.04$. This is consistent with its trajectory being carried by its accumulating normalization statistics rather than by its weight-update rule, though we have not tested that directly, and its curve should be read with that caveat. Macaque electrophysiology shows the same collapse at V1/V2, where the rule differences shrink toward the training resolution: directional cross-species support, from a single seed (see Limitations).

\subsection{What survives: a genuine learning effect at LOC}

\begin{figure*}[tp]
\centering
\includegraphics[width=\textwidth]{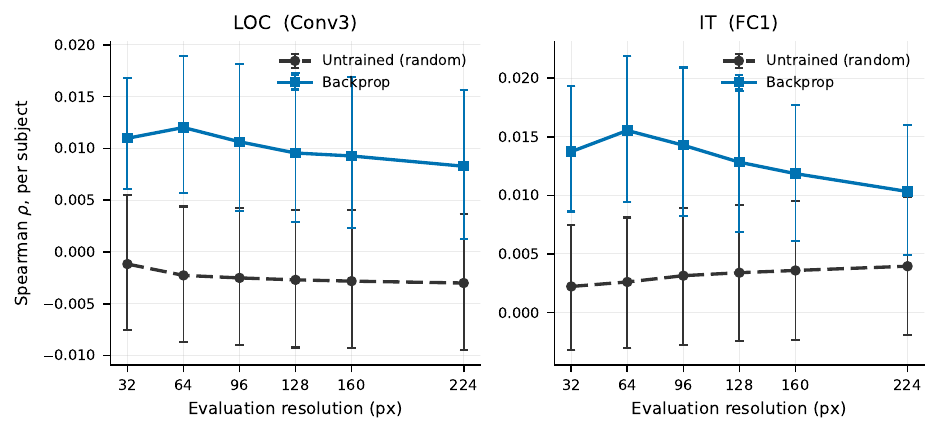}
\caption{\textbf{A genuine learning effect survives at higher areas.} Backprop vs.\ untrained alignment (per subject, cross-run pairs; 95\% stimulus-bootstrap CIs, $5$ seeds) across resolution at LOC (left) and IT (right). Backprop exceeds the untrained baseline at every resolution ($5/5$ seeds), unlike at V1.}
\label{fig:higher}
\end{figure*}

One effect holds up across the sweep (Fig.~\ref{fig:higher}). At LOC, backprop-trained networks align better than untrained ones at every resolution (backprop$-$untrained $= \NV{gap.LOC.32} \pm \NV{gap.LOC.32.sem}$ at $32$\,px and $\NV{gap.LOC.224} \pm \NV{gap.LOC.224.sem}$ at $224$\,px; $5/5$ seeds throughout). IT shows a weaker but consistently positive version of the same thing ($\NV{gap.IT.32} \pm \NV{gap.IT.32.sem}$ at $32$\,px, $\NV{gap.IT.224} \pm \NV{gap.IT.224.sem}$ at $224$\,px; $5/5$ seeds). What is stable across resolution here is the difference, not the levels: backprop's own LOC alignment falls from $\NV{rho.bp.LOC.32}$ to $\NV{rho.bp.LOC.224}$ over the sweep, but the untrained baseline sits near zero and slightly negative throughout ($\NV{rnd.LOC.max}$ to $\NV{rnd.LOC.min}$), so the sign and approximate size of the gap never change. Learning does leave a mark on the representations that this analysis choice does not erase, and it sits at higher areas rather than at V1.

\section{Discussion}

\textbf{What the result means.} One of the conclusions that has drawn interest to biologically plausible and untrained models, that they rival backpropagation at early visual cortex, is in this setting largely a function of the evaluation resolution. Work on biologically plausible rules is especially exposed to it, because those rules are what force the small-scale training that opens a gap between a model's native resolution and the resolution at which brain stimuli are modeled. A terminological note: strictly, evaluation resolution is an effect modifier rather than a confounder in the causal sense, since it is an analysis choice that changes the size and sign of the measured difference rather than a common cause of both. We use ``confound'' in the informal sense standard in this literature.

\textbf{Which resolution is ``correct''?} We are not claiming that $32$\,px is right and $224$\,px wrong, but the choice is less arbitrary than the practice suggests. A $3\times3$ Conv1 filter spans about $9\%$ of the image width at $32$\,px and about $1.3\%$ at $224$\,px, so evaluation resolution sets the effective receptive-field size of the model's early layers in units of the image, and thence in degrees of visual angle once the stimulus subtense during acquisition is known. Matching that to the receptive fields of the recorded neurons is a principled criterion, and it is applied elsewhere: \citet{laskar2018} choose their downsampling factor for exactly this reason. It is not applied in the comparisons this paper is about, which is what makes the choice look free.

Our data are consistent with such an account without establishing it. Section~3.3 shows the ResNet-50 and the Swin-Tiny peaking at low resolution despite being trained at $224$\,px, which refutes train/eval matching but is also what a receptive-field criterion predicts, since the filters of their early stages are small in image units at $224$\,px too. Section~3.5 fits it: in the upsampled arm the filters cover the same fraction of the image as in the native arm, and the gap appears in full as soon as the input exceeds the training resolution, whether or not image content above that resolution is present. What the account does not explain is why that content, where present, narrows the gap. Until such a criterion is adopted, the practical recommendation stands: if the ranking moves with an analysis choice, it cannot be read as a property of the learning rule until that choice is fixed, so evaluate at the training resolution and at several others, and state the resolution used.

\textbf{Mechanism.} None of four accounts explains the resolution dependence: train/eval resolution matching, low-level Gabor and pixel statistics, the normalization state of the untrained baseline, and convergence of the pooled descriptor toward a global brightness statistic. Three of the four are excluded by interventions that hold the convolutional weights bit-identical, which is the strongest form available here. What we can say positively is narrower than a mechanism but not nothing: limiting the image content to the training resolution leaves the V1 gap at $224$\,px intact ($R = \NV{s4.V1.R}$, \S3.5), so the gap follows the input size relative to the training resolution and is carried by a decline of backprop; image content above the training resolution narrows it. Why trained filters lose alignment once the input exceeds their training scale is the question that remains. The train/eval-matching account is the one to single out, because it is the explanation we ourselves flagged as a limitation in the endpoint study \citep{leutenegger2026}, on the reasoning that random filters are roughly scale-invariant while trained filters are tuned to $32$\,px statistics. The sweep turns that limitation into a quantified main effect and, at the same time, refutes the cause it proposed, since two networks trained at $224$\,px also peak at low resolution.

\subsection{A correction to our own earlier work}

The predictive-coding and STDP conditions here, in the endpoint study \citep{leutenegger2026}, and in its cross-species and training-dynamics companions share an implementation in which those two classes overrode \texttt{eval()} with a no-op, so their batch-normalization layers remained in training mode during feature extraction and their features were re-normalized by the statistics of each evaluation batch, while random, backprop and feedback alignment used stored running statistics. Repairing this changes the two affected conditions substantially. Of the other three, random is unchanged exactly and feedback alignment to within $10^{-5}$; backprop moves only within run-to-run reproducibility of the same pipeline, which at Conv1$\to$V1 is not tighter than about $2\times10^{-3}$, with a maximum of $5\times10^{-3}$ across all layer--ROI cells. The repair is therefore not detectable in backprop above the noise of re-running it. In particular, the flat resolution profile that made STDP look distinctive was an artifact: repaired, it declines with resolution like every other trained condition. Every number in this paper comes from the repaired implementation. Correction notes identifying the affected results accompany the current arXiv versions of all three earlier preprints. The training-dynamics result is the one that does not survive repair: with the defect fixed, predictive coding degrades V1 alignment more than backpropagation does rather than less, reversing that paper's central claim. The random and backprop conditions, which carry the untrained-versus-trained claim of the endpoint study, are unaffected.

\subsection{Relation to prior work}

Input resolution is not usually treated as a variable in this literature, but it is not ignored either: \citet{laskar2018} downsample their stimuli by a fixed factor specifically to match the effective receptive-field size of the model to that of the recorded neurons, and report that the choice changes their fits. Our contribution is to treat that choice as a variable rather than a setting, and to show that the conclusion moves with it. That work also contrast-normalizes its stimuli before model evaluation, which removes the luminance channel we find to be doing so much work here; whether the model--brain comparisons that do not normalize are measuring something different is a question our \S3.4 raises and does not settle. The study grew out of a limitation in our endpoint comparison \citep{leutenegger2026}, which reported the untrained $>$ backprop effect at V1, showed it survived a pixel-similarity partial-RSA control, and found the Conv1 filters to be color-opponent rather than Gabor-like, with no difference in Gabor-peakedness between rules. Those controls already pointed away from pure pixel similarity and from Gabor structure; the resolution sweep shows what was actually carrying the effect. Our own repaired sweep gives $\Delta\rho = \NV{gap.V1.224}$ at the same cell in the same convention (the endpoint study's revised value is $\NV{p1.gap}$). The untrained condition is bit-identical between the two studies, agreeing to $5\times10^{-7}$; the entire difference sits in backprop, which does not reproduce across pipelines. More generally, reports that untrained networks are brain-like at early visual cortex \citep{saxe2011,truzzi2025} are best read with the evaluation resolution in view, especially when a network is trained at one resolution and aligned at another.

\subsection{Limitations}

The claims above are bounded by the data in several ways. (1)~The macaque analysis is single-seed, so the phrase ``across two species'' should be read as directional support rather than as two independently powered replications; the human data carry the seed-level statistics throughout. (2)~The macaque datasets use different stimuli for different regions (FreemanZiemba textures vs.\ HVM objects), a known confound in that comparison. (3)~The human fMRI data comprises three subjects, one of which shows consistently weak signal, so the core findings rest on two of them. (4)~STDP at higher areas rests on a checkpoint whose FC1 and normalization layers are less reliable, so we keep strong STDP claims to the early layers. (5)~The conditions differ widely in how far training moves the weights. Predictive coding displaces its first convolutional layer by $2\%$ of the initialization norm and does not update its third layer at all, so at the layers mapped to V1 and LOC it is close to, or identical with, the untrained network; statements that a result holds across conditions should be read with that in mind (Appendix~B). (6)~Feedback alignment and STDP initialize their convolutional weights with \texttt{kaiming\_normal\_}, whereas the remaining conditions use PyTorch's default \texttt{kaiming\_uniform\_}; the resulting initialization scale differs by a factor of roughly $2.4$ at Conv1. The single untrained condition is therefore the matched control for three of the five conditions rather than all five, and each rule's own epoch-$0$ network provides the matched baseline where one is required. (7)~The joint partial RSA against four correlated low-level references drives every trained condition strongly negative, which is over-subtraction; the fraction of V1 alignment it removes should be read as an upper bound on what global image statistics could explain, not as a corrected estimate. (8)~That a scalar luminance value matches our best model at V1 bounds what any of these comparisons can resolve in this dataset, including ours; the finding is about the measurement, and stronger fMRI or electrophysiology may not share it. (9)~The dose--response argument rests on six resolutions evaluated on the same $720$ images with the same weights; these are six correlated measurements of one manipulation, not six independent tests, and the monotonicity should be read as the shape of a single effect rather than as replication. (10)~The luminance-convergence ordering rests on five conditions at $n=5$ seeds, and the within-filter test that separates it from V1 alignment turns on a single calibration variant; both are small samples, and we treat the ordering as an observation rather than as an explained relationship. (11)~All results use rank correlation between RDMs built from globally pooled features. That discards spatial structure before the comparison, so the statements about how much of V1 alignment a single brightness scalar accounts for are specific to that choice; a fitted readout on the full feature map might place the models higher, and we have not tested one. (12)~The $720$ evaluation stimuli are single-presentation trials, so the fMRI RDMs are built from single-trial rather than trial-averaged responses. Absolute $\rho$ values therefore carry measurement noise that repetition averaging would remove, and no within-subject reliability can be estimated on these stimuli. The relative comparisons that carry our claims are unaffected, but the absolute scale should not be read as a property of V1. (13)~The upsampled images of \S3.5 are smoother than natural images of the same size. The comparison separates information content from input size, but it does not isolate which image property is responsible for backprop's decline.

\subsection{Future work}

Several of the questions this study raises are answerable with the same infrastructure. Section~3.5 ties the V1 gap to the input size relative to the training resolution, which narrows the search without ending it: the next question is which property of the enlarged input lowers backprop's alignment, and why image content above the training resolution partly offsets it. A second is whether the effect survives a readout that does not discard spatial structure; our comparison rank-correlates globally pooled features, whereas a fitted readout on the full feature map is standard elsewhere and is the obvious robustness check on the scale results of \S3.4. A third is the receptive-field criterion discussed above, which requires the stimulus subtense during acquisition and a matched-angle sweep. Whether the phenomenon appears in architectures without batch normalization, and whether calibration procedures outside the family we constructed behave differently, are both open; so is why calibration costs any alignment at all, given that its cost does not track how closely the stored statistics match the evaluation activations. Swin-Tiny extends the result to one transformer family, and other ViT families remain to be tested rather than assumed. We have no positive mechanism, and we would rather leave that distinction visible than close it prematurely.

\section{Conclusion}

RSA comparisons of learning rules and architectures at early visual cortex are governed by an evaluation-resolution confound that can manufacture or substantially change the apparent advantage of untrained networks. It holds across conditions, species, training time, and three architecture families. Four candidate mechanisms do not explain it: train/eval resolution matching, low-level Gabor and pixel accounts, the normalization state of the untrained baseline, and convergence of the pooled descriptor toward a global brightness statistic, three of them excluded by interventions that hold the convolutional weights bit-identical. A fifth experiment, analysed against a criterion fixed before the analysis, locates it: with image content limited to the training resolution the V1 gap at $224$\,px is as large as with native content ($R = \NV{s4.V1.R}$ $[\NV{s4.V1.R.lo}, \NV{s4.V1.R.hi}]$), so it arises once the input exceeds the training resolution, driven by a decline of backprop, and content above that resolution narrows it rather than creating it. Along the way we find that a single scalar luminance value per image matches the untrained network's V1 alignment here, and that luminance similarity orders the conditions exactly as their resolution slopes do without carrying the effect. That is a caution about what this comparison can resolve at all. In this setting, the learning effect that holds across resolution sits at a higher area (LOC). For this line of work, evaluation resolution is something to control and to report.

\subsection*{Acknowledgements}
The author thanks the creators of the THINGS-fMRI, FreemanZiemba2013, and MajajHong2015 datasets and the Brain-Score team for their infrastructure. Analysis code and text drafts were developed with the assistance of an AI coding assistant; all results were verified by the author.

\subsection*{Code and Data Availability}
Code and results: \url{https://github.com/nilsleut/evaluation-resolution-rsa/tree/arxiv-v3}.

\subsection*{Appendix A: Hyperparameter Details}
\begingroup
\footnotesize
\begin{tabular}{@{}llp{3.3cm}@{}}
\toprule
Condition & lr & Notes \\
\midrule
BP & $10^{-3}$ (Adam) & weight decay $10^{-4}$, cosine schedule, gradient clip $1.0$, dropout $0.3$, cross-entropy \\
FA & $5\times10^{-4}$ (SGD) & momentum $0.9$, fixed random feedback weights \\
PC & $10^{-4}$ & $T=10$ inference steps, inference rate $0.02$; Adam-trained readout \\
STDP & $5\times10^{-4}$ & $A_\pm=0.003$, $\tau_\pm=20$\,ms, $10$ timesteps; Adam-trained readout \\
\bottomrule
\end{tabular}
\endgroup

\noindent All conditions share the architecture and training setup in Section~2 (three convolutional blocks $32/64/128$, FC1 $512$, FC2 $10$; $8{,}000$-image CIFAR-10 subset at $32\times32$, batch size $128$, $40$ epochs, $5$ seeds). fMRI RDMs were computed with correlation distance on single-trial BOLD responses. The $720$ evaluation stimuli are single-presentation trials in all three subjects, so no trial averaging occurs for any of them; the averaging step in the extraction code is a no-op here. No additional z-scoring or outlier removal was applied beyond the THINGS-fMRI preprocessing pipeline.

\subsection*{Appendix B: Weight Displacement from Initialization}

\noindent Relative change $\lVert W_{40} - W_{0}\rVert / \lVert W_{0}\rVert$ in the convolutional weights over $40$ epochs (seed $42$), measured against the shared initialization. Only backpropagation and predictive coding share that initialization; feedback alignment and STDP use a different scheme and are therefore not comparable on this baseline (Limitations, item 6).

\begingroup
\footnotesize
\begin{tabular}{@{}llrr@{}}
\toprule
Layer & $\to$ ROI & Backprop & Predictive Coding \\
\midrule
Conv1 & V1/V2 & $29.4\%$ & $2.1\%$ \\
Conv2 & --    & $90.4\%$ & $7.2\%$ \\
Conv3 & LOC   & $119.2\%$ & $0.0\%$ \\
\bottomrule
\end{tabular}
\endgroup

\subsection*{Appendix C: Best-Layer Selection}

\noindent Instead of the fixed Conv1$\to$V1 mapping, selecting for each condition and resolution whichever layer aligns best with V1. The best layer does not drift with resolution except for predictive coding, and the Random$-$Backprop gap grows in the same way.

\begingroup
\footnotesize
\begin{tabular}{@{}lccc@{}}
\toprule
 & $32$\,px & $96$\,px & $224$\,px \\
\midrule
Best layer, random & Conv2 & Conv2 & Conv2 \\
Best layer, backprop & Conv1 & Conv1 & Conv1 \\
Best layer, FA / STDP & Conv2 & Conv2 & Conv2 \\
Best layer, PC & Conv1 & Conv1 & Conv2 \\
\midrule
Gap, fixed Conv1 & $\NV{gap.V1.32}$ & $\NV{gap.V1.96}$ & $\NV{gap.V1.224}$ \\
Gap, best layer & $\NV{bestgap.32}$ & $\NV{bestgap.96}$ & $\NV{bestgap.224}$ \\
Seeds positive (best layer) & $\NV{bestgap.32.pos}/5$ & $\NV{bestgap.96.pos}/5$ & $\NV{bestgap.224.pos}/5$ \\
\bottomrule
\end{tabular}
\endgroup

\bibliographystyle{plainnat}

\begin{thebibliography}{99}

\bibitem[Bi and Poo(1998)]{bi1998}
Bi, G.-q. and Poo, M.-m. (1998). Synaptic modifications in cultured hippocampal neurons. \textit{J.\ Neurosci.}, 18:10464--10472.

\bibitem[Freeman et~al.(2013)]{freeman2013}
Freeman, J., Ziemba, C.~M., Heeger, D.~J., Simoncelli, E.~P., and Movshon, J.~A. (2013). A functional and perceptual signature of the second visual area in primates. \textit{Nature Neuroscience}, 16:974--981.

\bibitem[Goddard and Mullen(2020)]{goddard2020} Goddard, E. and Mullen, K.~T. (2020). fMRI representational similarity analysis reveals graded preferences for chromatic and achromatic stimulus contrast across human visual cortex. \textit{NeuroImage}, 215:116780.

\bibitem[He et~al.(2016)]{he2016}
He, K., Zhang, X., Ren, S., and Sun, J. (2016). Deep residual learning for image recognition. \textit{CVPR}, pp.~770--778.

\bibitem[Hebart et~al.(2023)]{hebart2023}
Hebart, M.~N., Contier, O., Teichmann, L., Rockter, A.~H., Zheng, C.~Y., Kidder, A., Corriveau, A., Vaziri-Pashkam, M., and Baker, C.~I. (2023). THINGS-data, a multimodal collection of large-scale datasets for investigating object representations in human brain and behavior. \textit{eLife}, 12:e82580.

\bibitem[Kriegeskorte et~al.(2008)]{kriegeskorte2008}
Kriegeskorte, N., Mur, M., and Bandettini, P. (2008). Representational similarity analysis: connecting the branches of systems neuroscience. \textit{Frontiers in Systems Neuroscience}, 2:4.

\bibitem[Laskar et~al.(2018)]{laskar2018} Laskar, M.~N.~U., Sanchez Giraldo, L.~G., and Schwartz, O. (2018). Correspondence of deep neural networks and the brain for visual textures. \textit{arXiv:1806.02888}.

\bibitem[Leutenegger(2026)]{leutenegger2026}
Leutenegger, N. (2026). Untrained CNNs exceed backpropagation in V1 alignment at high evaluation resolution: A systematic RSA comparison of four learning rules against human fMRI. \textit{arXiv:2604.16875}, v4.

\bibitem[Nili et~al.(2014)]{nili2014}
Nili, H., Wingfield, C., Walther, A., Su, L., Marslen-Wilson, W., and Kriegeskorte, N. (2014). A toolbox for representational similarity analysis. \textit{PLoS Computational Biology}, 10:e1003553.

\bibitem[Lillicrap et~al.(2016)]{lillicrap2016}
Lillicrap, T.~P., Cownden, D., Tweed, D.~B., and Akerman, C.~J. (2016). Random synaptic feedback weights support error backpropagation for deep learning. \textit{Nature Communications}, 7:13276.

\bibitem[Liu et~al.(2021)]{liu2021}
Liu, Z., Lin, Y., Cao, Y., Hu, H., Wei, Y., Zhang, Z., Lin, S., and Guo, B. (2021). Swin Transformer: Hierarchical vision transformer using shifted windows. \textit{ICCV}, pp.~10012--10022.

\bibitem[Majaj et~al.(2015)]{majaj2015}
Majaj, N.~J., Hong, H., Solomon, E.~A., and DiCarlo, J.~J. (2015). Simple learned weighted sums of inferior temporal neuronal firing rates accurately predict human core object recognition performance. \textit{J.\ Neurosci.}, 35:13402--13418.

\bibitem[Masquelier and Thorpe(2007)]{masquelier2007}
Masquelier, T. and Thorpe, S.~J. (2007). Unsupervised learning of visual features through spike timing dependent plasticity. \textit{PLOS Computational Biology}, 3(2):e31.

\bibitem[Olshausen and Field(1996)]{olshausen1996}
Olshausen, B.~A. and Field, D.~J. (1996). Emergence of simple-cell receptive field properties by learning a sparse code for natural images. \textit{Nature}, 381:607--609.

\bibitem[Rao and Ballard(1999)]{rao1999}
Rao, R.~P.~N. and Ballard, D.~H. (1999). Predictive coding in the visual cortex. \textit{Nature Neuroscience}, 2:79--87.

\bibitem[Saxe et~al.(2011)]{saxe2011}
Saxe, A.~M., Koh, P.~W., Chen, Z., Bhand, M., Suresh, B., and Ng, A.~Y. (2011). On random weights and unsupervised feature learning. \textit{ICML}.

\bibitem[Schrimpf et~al.(2020)]{schrimpf2020}
Schrimpf, M., Kubilius, J., Hong, H., et~al. (2020). Brain-Score: Which artificial neural network for object recognition is most brain-like? \textit{bioRxiv}.

\bibitem[Truzzi and Cusack(2025)]{truzzi2025}
Truzzi, A. and Cusack, R. (2025). Neural responses in early visual cortex are well predicted by random-weight CNNs. \textit{bioRxiv}.

\bibitem[Whittington and Bogacz(2017)]{whittington2017}
Whittington, J.~C.~R. and Bogacz, R. (2017). An approximation of the error backpropagation algorithm in a predictive coding network with local Hebbian synaptic plasticity. \textit{Neural Computation}, 29:1229--1262.

\bibitem[Yamins and DiCarlo(2016)]{yamins2016}
Yamins, D.~L.~K. and DiCarlo, J.~J. (2016). Using goal-driven deep learning models to understand sensory cortex. \textit{Nature Neuroscience}, 19:356--365.

\end{thebibliography}

\end{document}